\documentstyle[prl,aps]{revtex}
\begin{document}


\vskip 4mm

\centerline{\large \bf Isotope Effect in the Presence of Magnetic and
Nonmagnetic Impurities}

\vskip 4mm

\centerline{L. A. Openov and I. A. Semenihin}

\vskip 2mm

\centerline{\it Moscow State Engineering Physics Institute (Technical
University), 115409 Moscow, Russia}

\vskip 2mm

\centerline{R. Kishore}

\vskip 2mm

\centerline{\it Insituto Nacional de Pesquisas Espaciais, C.P. 515,
S. J. Campos, S. P., 12201-970, Brasil}

\vskip 6mm

\begin{quotation}

The effect of impurities on the isotope coefficient is studied theoretically
in the framework of Abrikosov-Gor'kov approach generalized to account for
both potential and spin-flip scattering in anisotropic superconductors. An
expression for the isotope coefficient as a function of the critical
temperature is obtained for a superconductor with an arbitrary contribution
of spin-flip processes to the total scattering rate and an arbitrary degree
of anisotropy of the superconducting order parameter, ranging from isotropic
$s$ wave to $d$ wave and including anisotropic $s$ wave and mixed
($s$+$d$) wave as particular cases. It is found that both magnetic
and nonmagnetic impurities enhance the isotope coefficient, the enhancement
due to magnetic impurities being generally greater than that due to
nonmagnetic impurities. From the analysis of the experimental results on LSCO
high temperature superconductor, it is concluded that the symmetry of the
pairing state in this system differs from a pure $d$ wave.

\end{quotation}

\vskip 6mm

The isotope effect has played an important role in the development of phonon
mediated mechanism of electron pairing in superconductors. Its
discovery gave rise to the
theory of phonon mediated mechanism of electron pairing by Bardeen, Cooper
and Schrieffer \cite{BCS}. The BCS theory  gave the isotope coefficient
$\alpha$, defined by the relation $T_{c}\propto M^{-\alpha}$,
or equivalently,
$\alpha = -\frac{\partial \ln T_{c}}{\partial \ln M}$,
equal to 1/2, in agreement with the experiments for mercury
\cite{max}. For simple superconducting metals like Hg, Zn, S, Pb,
etc., the values of $\alpha$ were found to be very close to the BCS value
1/2. The deviations from the BCS theory found in superconducting transition
metals and their compounds were reasonably well explained by taking into
account the effects of Coulomb interactions \cite{morel},
left out in the BCS theory, and by more realistic treatments based on
Eliashberg equations \cite{carbotte}.

The discovery of high temperature superconductors (HTSCs) brought a serious
challenge to the BCS theory which had firmly established the phonon mediated
mechanism of electron pairing as a dominant cause of superconductivity in
most previously known superconductors. For conventional superconductors, both
theory and experiment agree with the fact that $\alpha$ approaches the BCS
value of 1/2 as $T_{c}$ becomes larger. However, this trend is violated in
copper oxide HTSCs, where $\alpha$ has been found to vary with doping in
different ways \cite{Franck}. For the optimum doping, corresponding to the
highest critical temperature, the value of $\alpha$ is usually quite small
compared to the BCS value of 1/2. For a certain doping level, e.g. in some
La$_2$CuO$_4$ based HTSCs, the value of $\alpha$ becomes greater than 1/2
\cite{Franck}. Such a behavior of $\alpha$ has been considered as one of the
strongest evidence for a new mechanism of superconductivity in HTSCs.
However, this argument is not as strong as it appears. For example, it was
shown that, within the BCS picture or within the more realistic Eliashberg
approach, these deviations can be understood \cite{Kishore} by considering
the effects of anharmonicity, energy dependence of the electronic density of
states, pair breaking effects, isotope mass dependence of the carrier concentration,
etc.

The presence of impurities strongly affects various characteristics of HTSCs,
including the isotope
coefficient \cite{soerensen,babushkina,morris}. Earlier theoretical attempts
to describe the isotope effect in impure HTSCs were based on either the
Abrikosov-Gor'kov formula for $T_c$ of an isotropic $s$-wave superconductor
containing magnetic impurities \cite{babushkina,kresin1} or the
Abrikosov-Gor'kov-like formula for $T_c$ of an anisotropic superconductor
that contains nonmagnetic impurities only \cite{buzea,scattoni}. An
increase in $\alpha$ with impurity concentration has been demonstrated,
in qualitative agreement with the experiment. However, the theory predicted
the universal dependence of the normalized isotope coefficient,
$\alpha/\alpha_0$, on the normalized critical temperature, $T_c/T_{c0}$,
where $\alpha_0$ and $T_{c0}$ are, respectively, the values of $\alpha$ and
$T_c$ in the absence of impurities. This prediction does not agree with the
experimental findings \cite{soerensen,babushkina,morris}.
It was shown by Kresin {\it et al.} \cite{kresin1} that the universal
behavior of $\alpha/\alpha_0$ versus $T_c/T_{c0}$ is restricted to the case
that magnetic impurities are the only cause for the increase of
$\alpha/\alpha_0$,
and that other effects like the nonadiabaticity of the apex oxygen in
YBa$_2$Cu$_3$O$_7$ could break the universality. Here we provide an
alternative explanation of how the nonuniversality of the dependence of
$\alpha/\alpha_0$ on $T_c/T_{c0}$ arises. Our approach is based on taking
into account the combined effect of both nonmagnetic and magnetic
scatterers on the critical temperature of a superconductor with anisotropic
superconducting order parameter.

To derive the expression for the isotope coefficient, we make use of the
equation for $T_c$ of an anisotropic superconductor
containing both nonmagnetic and magnetic impurities \cite{openov}:
\begin{equation}
\ln\left(\frac{T_{c0}}{T_c}\right)=(1-\chi)\left[\Psi\left(\frac{1}{2}+%
\frac{1}{2\pi T_c\tau_m^{ex}}\right)-\Psi\left(\frac{1}{2}\right)\right]+%
\chi\left[\Psi\left(\frac{1}{2}+\frac{1}{4\pi T_c\tau}\right)-%
\Psi\left(\frac{1}{2}\right)\right],
\label{Tc}
\end{equation}
where $\Psi$ is the digamma function; $\tau_m^{ex}$ is the electron
relaxation time due to exchange scattering by magnetic impurities; $\tau$ is
the total electron relaxation time due to potential scattering by both
magnetic and nonmagnetic impurities, as well as due to exchange scattering by
magnetic impurities.
Eq. (\ref{Tc}) was obtained in Ref. \cite{openov} within the weak-coupling
limit of the BCS model in the framework of Abrikosov-Gor'kov approach.
It generalizes the well-known expressions \cite{Abrikosov1,Abrikosov2} for
the critical temperature of impure superconductors to the case of combined
effect of both nonmagnetic and magnetic impurities on the critical
temperature of anisotropic superconductors.
The coefficient
$\chi=1-\frac{\langle\Delta({\bf p})\rangle^2_{FS}}%
{\langle\Delta^2({\bf p})\rangle_{FS}}$
quantifies the degree of
anisotropy of the order parameter $\Delta({\bf p})$ on the Fermi surface (FS),
where the angular brackets $\langle ... \rangle_{FS}$ stand for a FS average.
For isotropic $s$-wave pairing, $\langle\Delta({\bf p})\rangle^2_{FS}%
=\langle\Delta^2({\bf p})\rangle_{FS}$, and hence $\chi=0$. For a
superconductor with $d$-wave pairing we have $\chi=1$ since
$\langle\Delta({\bf p})\rangle_{FS}=0$. The range $0<\chi<1$ corresponds to
anisotropic $s$-wave or mixed $(d+s)$-wave pairing. The higher the
anisotropy of $\Delta({\bf p})$ (e.g., the greater the partial
weight of a $d$-wave in the case of mixed pairing), the closer to unity is
the value of $\chi$. Note that in two particular cases of ($i$) magnetic
scattering in an isotropic $s$-wave superconductor ($\chi =0$) and ($ii$)
nonmagnetic scattering only in a superconductor with arbitrary in-plane
anisotropy of $\Delta({\bf p})$ ($1/\tau_m^{ex}=0, 0\leq\chi\leq 1$),
Eq. (\ref{Tc}) reduces to the well-known expressions
\cite{Abrikosov1,Abrikosov2}.

It is convenient to specify the relative contribution of spin-flip
scattering rate, $1/\tau_m^{ex}$, to the total scattering rate, $1/\tau$, by
the dimensionless parameter $\gamma$ defined as
$\frac{1}{\tau_m^{ex}}=\gamma\frac{1}{\tau}$.
The greater is the relative contribution from exchange scattering by magnetic
impurities to $1/\tau$, the higher is the value of $\gamma$ ($\gamma$ ranges
from 0 in the absence of exchange scattering to 1 in the absence of
non-spin-flip scattering). In general, the value of $\gamma$ depends on the
scattering strengths of individual nonmagnetic and magnetic impurities, as
well as on their concentrations. At relatively low doping level, one can
expect $\gamma$ to depend only on the type of the host material and doping
elements, not on the impurity concentration \cite{openov}.

Differentiating Eq. (\ref{Tc}) for $T_c$ with respect to the isotopic mass
$M$ under a reasonable assumption that electron relaxation times and the
anisotropy coefficient $\chi$ do not depend on $M$, taking the definition of
the parameter $\gamma$ into account and using the definition of the isotope
coefficient $\alpha$, one has
\begin{equation}
\frac{\alpha}{\alpha_0} = \left[1-
(1-\chi)\frac{\gamma}{2\pi T_c\tau}%
\Psi^{\prime}\left(\frac{1}{2}+\frac{\gamma}{2\pi T_c\tau}\right)-%
\chi\frac{1}{4\pi T_c\tau}%
\Psi^{\prime}\left(\frac{1}{2}+\frac{1}{4\pi T_c\tau}\right)\right]^{-1}.
\label{alpha/alpha_0}
\end{equation}
Equation (\ref{alpha/alpha_0}) is obviously more general than equations
used previously for the analysis of the pair breaking effect on the isotope
coefficient in HTSCs \cite{babushkina,kresin1,buzea,scattoni}.
Indeed, Eq. (\ref{alpha/alpha_0}) does not rely on particular assumptions
about the symmetry of the superconducting state and the nature of impurities
({\it either} magnetic {\it or} nonmagnetic). This equation can be used for
an impure superconductor with {\it arbitrary} anisotropy of the
superconducting order parameter and {\it arbitrary} relative contributions of
potential and spin-flip scattering to the electron relaxation time. Such an
approach is of particular importance for HTSCs doped with various chemical
elements since, first, there is a strong evidence for a dominant $d$-wave
(i.e., highly anisotropic) order parameter in HTSCs \cite{tsuei2} with a
subdominant $s$-wave component \cite{schurrer}, and, second, a
lot of experiments give evidence for the presence of magnetic scatterers
(along with nonmagnetic ones) in doped HTSCs
\cite{Xiao,Mahajan,Williams,Awana}.

At given values of $\chi$ and $\gamma$, Eqs. (\ref{Tc}) and
(\ref{alpha/alpha_0}) define the dependence of $\alpha/\alpha_0$ on
$T_c/T_{c0}$. One can see from
Eqs. (\ref{Tc}) and (\ref{alpha/alpha_0}) that the dependence of
$\alpha/\alpha_0$ on $T_c/T_{c0}$ has a universal shape for both an isotropic
$s$-wave superconductor ($\chi=0$) with nonzero contribution of exchange
scattering to the total scattering rate ($0<\gamma\le 1$) and a $d$-wave
superconductor ($\chi=1$) with an arbitrary ratio of spin-flip and potential
scattering rates ($0\le\gamma\le 1$), as it is seen from Fig. 1.

Quite a different picture is realized in the case $0<\chi<1$, i.e., for a
mixed $(d+s)$-wave or an anisotropic $s$-wave superconductor.
In this case the behaviour of $\alpha/\alpha_0$ as a
function of $T_c/T_{c0}$ essentially depends on the value of $\gamma$. As
$T_c/T_{c0}$ goes to zero, i.e., in dirty superconductors, the value of
$\alpha/\alpha_0$ tends to $1/(1-\chi)$ in the absence of exchange scattering
($\gamma=0$), while $\alpha/\alpha_0$ grows as
$\alpha/\alpha_0 \propto T_{c0}/T_c$ in the case $\gamma\ne 0$ for all values
of $\chi$. Thus, the universality of $\alpha/\alpha_0$ versus $T_c/T_{c0}$
curve breaks down for $0<\chi<1$. This is consistent with experimental
observations. Indeed, the studies of isotope effect in impure HTSCs
\cite{soerensen,babushkina,morris} show that the curves of $\alpha/\alpha_0$
versus $T_c/T_{c0}$ vary with the type of impurities, i.e., with the value of
$\gamma$ (since different chemical elements contribute differently to
spin-flip and potential scattering rates of charge carriers). So, our results
seem to be in a qualitative agreement with the experiment if one assumes that
the superconducting state in HTSCs differs from a pure $s$ or $d$-wave, i.e.,
$\chi\ne 0$ and $\chi\ne 1$, as is also indicated by several experimental
observations \cite{schurrer,klemm}.

However, our theoretical consideration doesn't account for a
number of factors that could influence the isotope coefficient in impure
superconductors, e.g., the energy dependent electronic density of states,
nonadiabaticity, anharmonicity, etc. In particular,
the change in the carrier concentration $n_h$ upon chemical substitution can
have a strong influence on $T_c$ (and hence on $\alpha$) along with the
pair-breaking effect. This is why, to compare our calculations with the
experiment, we have taken the data on the isotope effect in the system
La$_{1.85}$Sr$_{0.15}$Cu$_{1-x}$M$_x$O$_4$ (M = Ni, Zn, Co, Fe), see
Ref. \cite{babushkina}. In this system, the critical temperature decreases
rapidly as the impurity content $x$ increases, and falls down to
$\approx 0.3T_{c0}$ already at $x=0.02-0.03$.
At such low impurity concentration, the change in $T_c$ due to
the change in $n_h$ can be neglected in the first approximation (as compared
with the pair-breaking effect) since the value of $T_{c0}\approx 40$K in the
impurity-free HTSC La$_{1.85}$Sr$_{0.15}$CuO$_4$ corresponds to the maximum
on the curve $T_c(n_h)$, and hence changes insignificantly with $n_h$ as long
as the change in $n_h$ is small. Note that such an approach (i.e., ignoring
the change in the carrier concentration upon doping) may not be appropriate
in case of YBa$_2$(Cu$_{1-x}$Zn$_x$)$_3$O$_7$ where $T_c/T_{c0}\approx 0.3$
at $x=0.06$ and all the more for Y$_{1-x}$Pr$_x$Ba$_2$Cu$_3$O$_7$ where
$T_c/T_{c0}\approx 0.3$ at $x=0.5$, see Ref. \cite{soerensen}.
A reasonable explanation of the isotope effect in these HTSCs has been given
by Kresin {\it et al.} \cite{kresin1} who considered the interplay between
the changes in the carrier concentration and nonadiabaticity of the apex
oxygen.

Fig.2 shows the experimental data along with theoretical graphs calculated
for $\chi=0.5$ and different values of $\gamma$, ranging from 0 to 1. One can
see that at low impurity content $x<0.008$ ($T_c/T_{c0}>0.75$) there is a
good agreement between the theory and the experiment. However, at higher
values of $x$ (i.e., at lower values of $T_c/T_{c0}$) the experimental points
lie well above the theoretical curves for all impurity elements, except for
M = Ni. The same is true for other values of $\chi$ since the upper curve
in Fig. 2 (for $\gamma=1$) changes insignificantly with $\chi$, at least for
$T_c/T_{c0}>0.2$.

To bring the theory closer to the agreement with the experiment, one can
assume that the value of $T_{c0}$ in the impurity-free HTSC
La$_{1.85}$Sr$_{0.15}$CuO$_4$ is strongly depressed relative to its
"intrinsic" value \cite{Kresin2} because of the scattering of charge
carriers by inhomogenities produced by substitution of Sr for La. Recently it
was suggested  \cite {baskaran} that the anomalous response of the
anisotropic superconducting
state to the development of low energy dynamical charge stripes
can also cause the suppression of the "intrinsic" $T_{c0}$. Taking $T_{c0}$
to be equal to its "intrinsic" value, it is possible to reach a qualitative
agreement with the experiment on the isotope effect in impure
HTSCs even for heavily doped samples with low $T_c$. Since the value of
"intrinsic" $T_{c0}$ in La$_{1.85}$Sr$_{0.15}$CuO$_4$ is not known
{\it a priori}, we take the suggested value \cite{baskaran} of 90 K.

Fig. 3 shows the results for the set of $T_{c0}=90$K and $\chi=0.5$. For such
a choice of the values of $T_{c0}$ and $\chi$, the theoretical curves for
$\gamma=0-1$ are closer to the experimental data \cite{babushkina}. Since
the value of $\alpha_0$ at "intrinsic" $T_{c0}$ is not known, we assumed for
simplicity that the value of $\alpha$ is the same at $T_{c0}\approx$ 40 K and
90 K. This assumption can be the reason for a discrepancy between the
theory and the experiment in the region $T_c/T_{c0}=0.3-0.4$, see Fig. 3,
since one could expect the value of $\alpha$ at $T_{c0}=90$K to be somewhat
lower than at $T_{c0}\approx 40$K. It should be noted that in our theory
different types of impurities correspond to different values of $\gamma$. The
magnetic elements Fe and Co can be thought of being characterized by
$\gamma$ in the range 0.1 - 1, while Ni, whose magnetic moment is reduced
considerably by doping \cite{babushkina}, can be assigned somewhat lower
value of $\gamma\approx 0.01$, see Fig. 3. Similiarly, the doping by the
nonmagnetic element Zn induces magnetic moments
\cite{Xiao,acquarone}. One can see from Fig. 3 that Zn can be
assigned the value of $\gamma$ in the range 0.01 - 0.05, i.e. lower than in
the case of Fe and Co but greater than in the case of Ni.

Finally, we note that the agreement between the theory and the experiment
becomes worse if one takes the value of $\chi$ closer to unity. This implies
that the symmetry of the pairing state in La-based HTSCs can differ
considerably from a pure {\it d}-wave. As far as we know, up to now there
were only indirect arguments in favor of {\it d}-wave symmetry of the
superconducting state in La$_{1.85}$Sr$_{0.15}$CuO$_4$ \cite{Takanaka}, while
phase-sensitive experiments are unknown to us. The value of $\chi\approx 0.5$
is however close to that expected for the two-dimensional order parameter of
the type $\Delta ({\bf k})$ = $\Delta_0$ [cos($k_x a$)+cos($k_y a$)],
see Ref. \cite{Openov2}.

In conclusion, we have studied theoretically the effect of both magnetic and
nonmagnetic impurities on the isotope coefficient in the framework of a
generalized Abrikosov-Gorkov approach for the anisotropic superconductors. We
have shown how the interplay between the potential and spin-flip impurity
scattering gives rise to the nonuniversal dependence of $\alpha/\alpha_0$
versus $T_{c}/T_{c0}$ in mixed $(d+s)$ wave or anisotropic $s$ wave
superconductors. Our main result is that if the impurities are viewed as the
only cause for the increase in the isotope coefficient in
La$_{1.85}$Sr$_{0.15}$Cu$_{1-x}$M$_x$O$_4$, then the symmetry of the
superconducting order parameter in La$_{1.85}$Sr$_{0.15}$CuO$_4$ appears to
be different from a pure $d$-wave.
Indeed, even if one takes into account relatively large
experimental errors in the values of $\alpha$, it is clearly seen that
experimental points do not lie on a single "universal" curve of
$\alpha/\alpha_0$ versus $T_c/T_{c0}$ as one could expect in the case of a pure
$d$-wave symmetry of the superconducting state. According to our calculations,
the difference in $\alpha/\alpha_0$ versus $T_{c0}/T_c$ curves for different
impurity elements can be attributed to different contributions from the
exchange scattering to the total scattering rate of charge carriers in the
mixed $(d+s)$-wave or anisotropic $s$-wave superconducting state. The agreement
with the experiment is much more better if one assumes that the "intrinsic"
value of $T_{c0}$ in La$_{1.85}$Sr$_{0.15}$CuO$_4$ reaches $\approx 90$ K, more
than twice greater than the experimental value $\approx 40$ K.
It would be interesting to check if it is possible to explain the
nonuniversality of $\alpha/\alpha_0$ versus $T_c/T_{c0}$ in
La$_{1.85}$Sr$_{0.15}$Cu$_{1-x}$M$_x$O$_4$ within a pure $d$-wave framework.

\vskip 2mm

L.A.O and I.A.S acknowledge the support from the Russian Federal Program
"Integration", projects No A0133 and No A0155. A part of the work was carried
out during the stay of R. K. at ICTP, Trieste, Italy as a Research Associate.

\newpage

\centerline{\bf FIGURE CAPTIONS}

\vskip 2mm

Fig. 1. Universal dependence of the normalized isotope coefficient
$\alpha/\alpha_0$ on the normalized critical temperature $T_c/T_{c0}$ in an
impure isotropic $s$-wave superconductor ($\chi=0$) with a finite
concentration of magnetic scatterers and an impure $d$-wave superconductor
($\chi=1$) with an arbitrary ratio of spin-flip and potential scattering
rates.

Fig. 2. Same as in Fig. 1 for $\chi=0.5$ (a specific case of anisotropic
pairing) for different values of the coefficient $\gamma$ specifying the
relative contribution to the total scattering rate from exchange scattering.
$\gamma=0$ (dot-dashed curve), 0.01 (thin solid curve), 0.05 (dashed curve),
0.1 (dotted curve), 1 (thick solid curve). Experimental data from
Ref. \cite{babushkina} for isotope effect in
La$_{1.85}$Sr$_{0.15}$Cu$_{1-x}$M$_x$O$_4$ with different
$x$ and M = Ni (triangles), Zn (open squares); Co (closed circles);
Fe (closed squares). Experimental values of $T_c$ as a function of $x$ are
normalized to the value of $T_{c0}$ = 37.5 K at $x=0$.

Fig. 3. Same as in Fig. 2 for $T_{c0}$ = 90 K, see text for details.

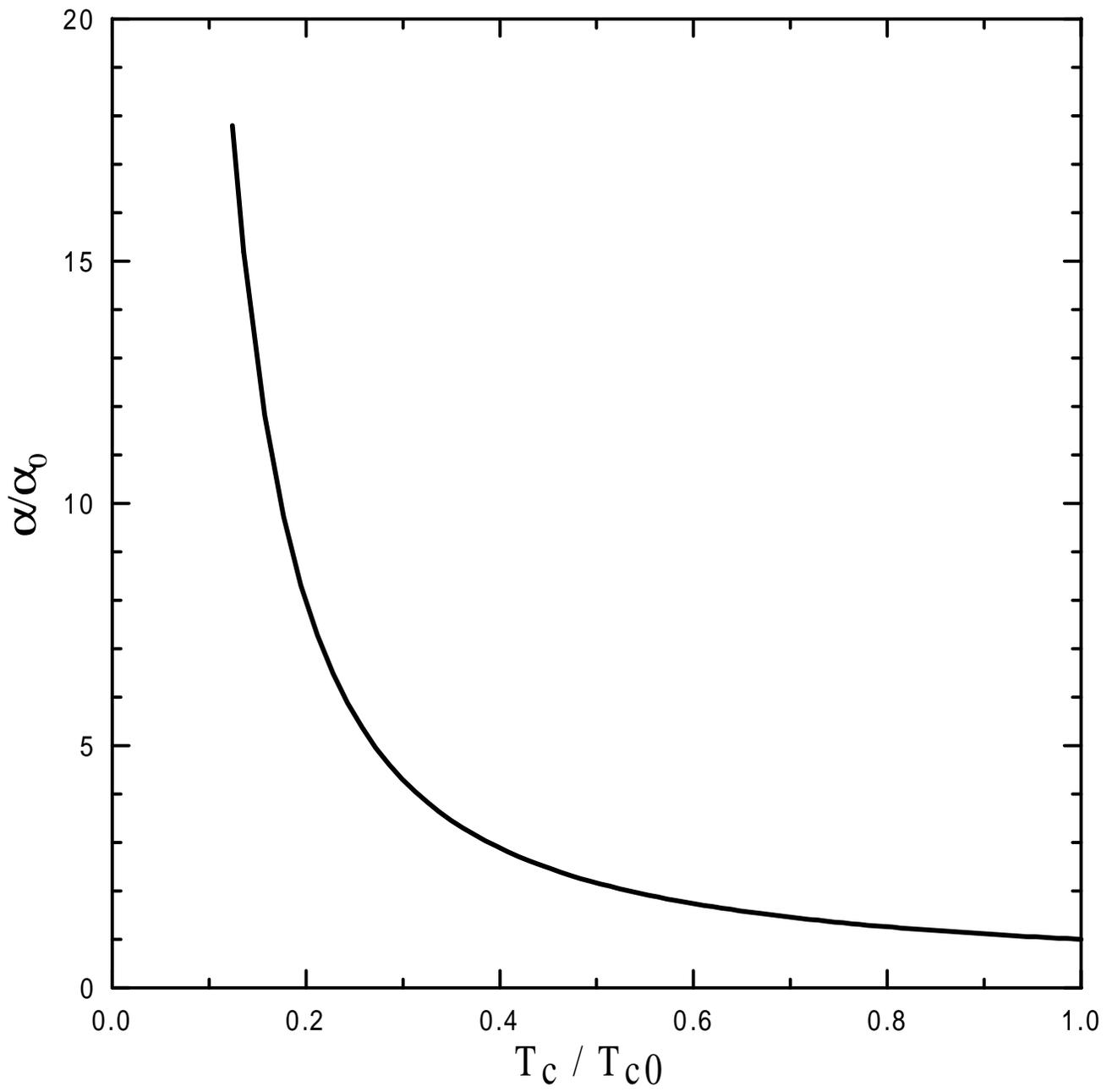
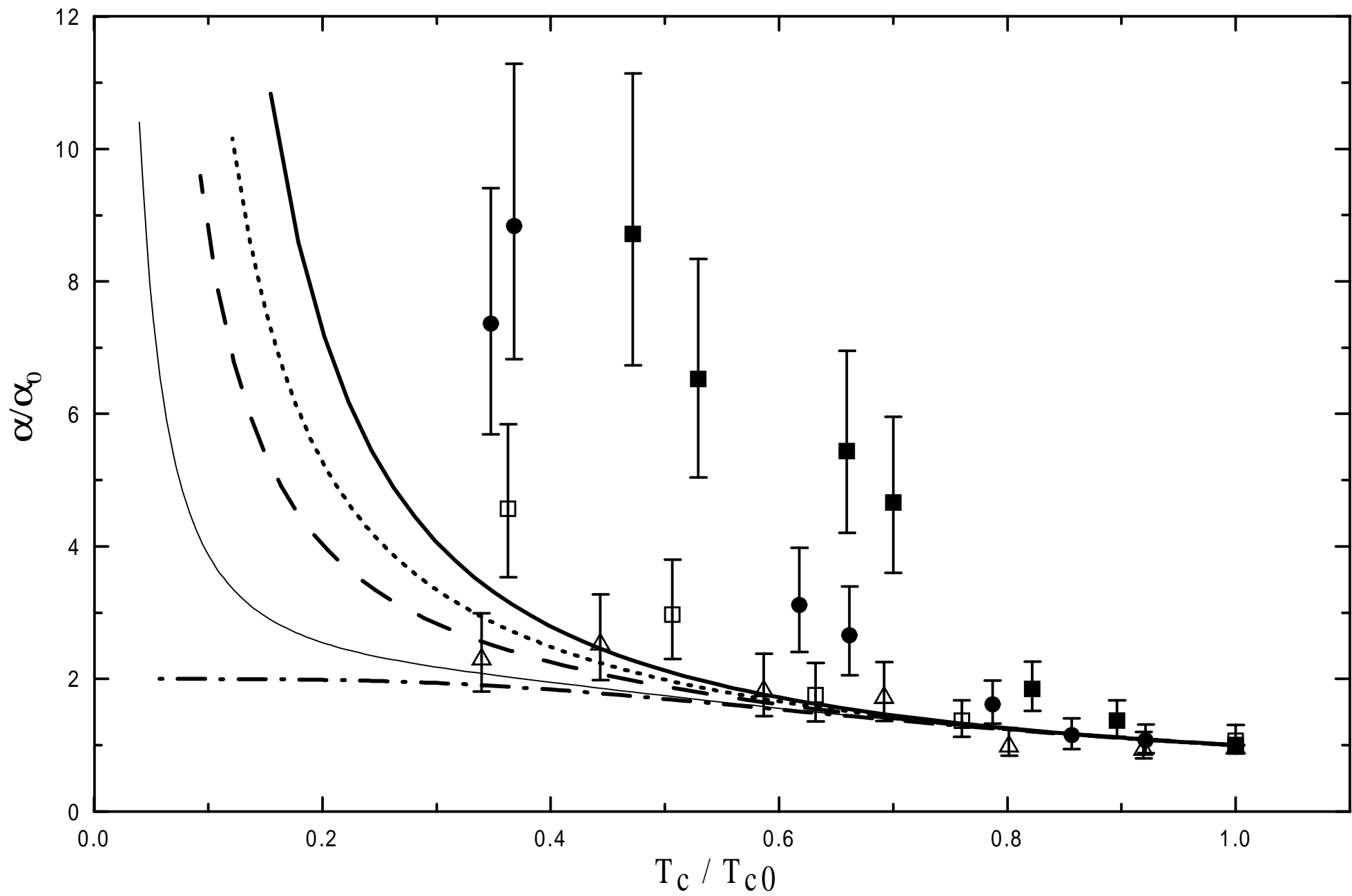
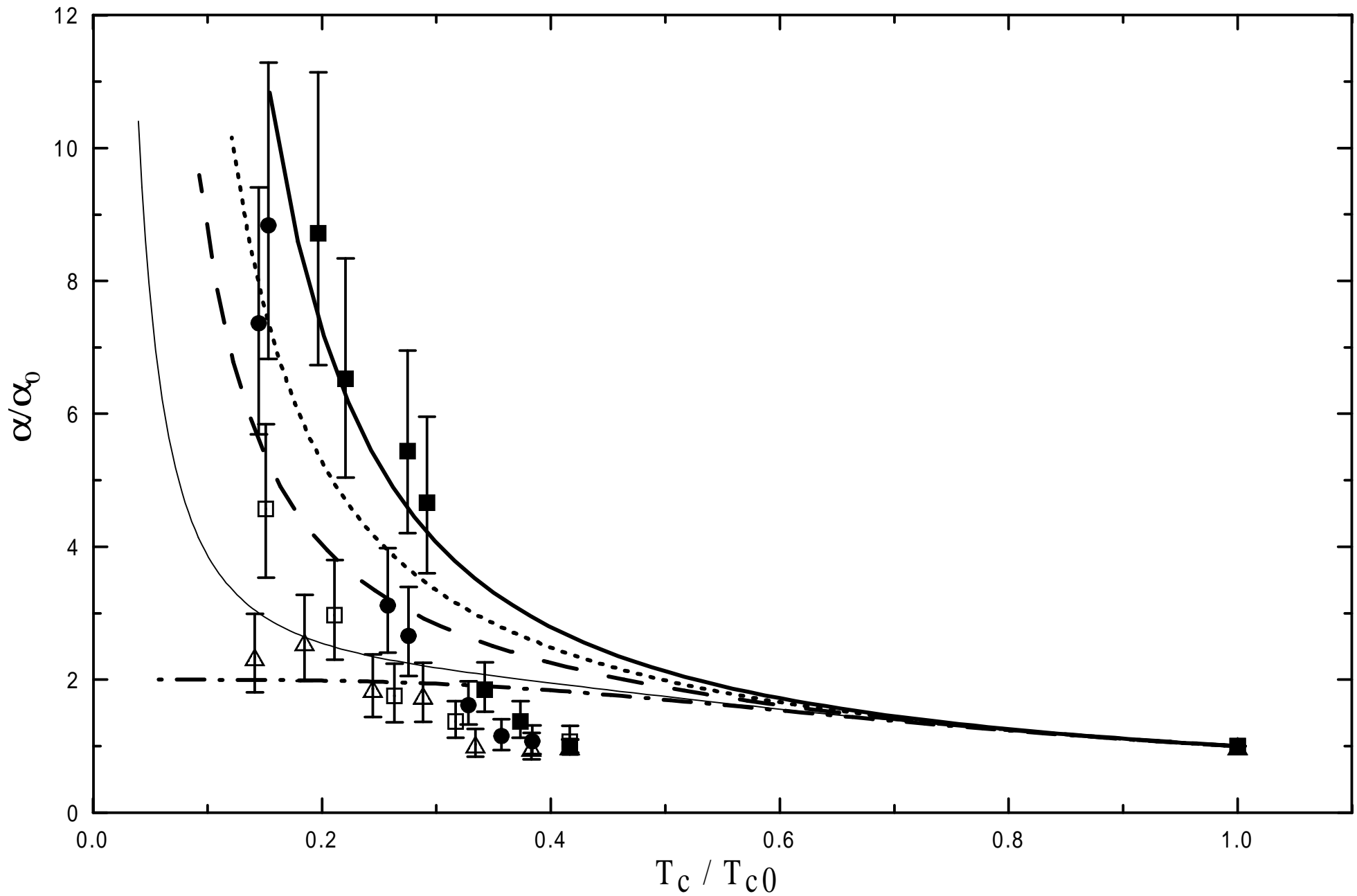
\end{document}